\documentclass[letterpaper,twocolumn,prl,aps,superscriptaddress,floatfix]{revtex4}
\usepackage{mathptmx}

\usepackage[latin9]{inputenc}
\usepackage{amsmath}
\usepackage{amssymb}
\usepackage{graphicx}
\usepackage{esint}
\usepackage{array}

\usepackage[unicode=true,pdfusetitle,
 bookmarks=true,bookmarksnumbered=false,bookmarksopen=false,
 breaklinks=false,pdfborder={0 0 0},pdfborderstyle={},backref=false,colorlinks=true]
 {hyperref}
\hypersetup{
 colorlinks,linkcolor=red,citecolor=blue}

\makeatletter

\pdfpageheight\paperheight
\pdfpagewidth\paperwidth

\@ifundefined{textcolor}{}
{%
 \definecolor{BLACK}{gray}{0}
 \definecolor{WHITE}{gray}{1}
 \definecolor{RED}{rgb}{1,0,0}
 \definecolor{GREEN}{rgb}{0,1,0}
 \definecolor{BLUE}{rgb}{0,0,1}
 \definecolor{CYAN}{cmyk}{1,0,0,0}
 \definecolor{MAGENTA}{cmyk}{0,1,0,0}
 \definecolor{YELLOW}{cmyk}{0,0,1,0}
}

\usepackage{xcolor}\usepackage{soul}

\newcommand{\bra}[1]{\ensuremath{\left\langle#1\right|}}
\newcommand{\ket}[1]{\ensuremath{\left|#1\right\rangle}}
\definecolor{blue}{rgb}{0,0,1}
\definecolor{red}{rgb}{1,0,0}
\definecolor{green}{rgb}{0,1,0}

\makeatother

\begin{document}
\title{Demonstration of Controlled-Phase Gates between Two Error-Correctable Photonic Qubits}
\author{Y.~Xu}
\thanks{These two authors contributed equally to this work.}
\affiliation{Center for Quantum Information, Institute for Interdisciplinary Information
Sciences, Tsinghua University, Beijing 100084, China}
\author{Y.~Ma}
\thanks{These two authors contributed equally to this work.}
\affiliation{Center for Quantum Information, Institute for Interdisciplinary Information
Sciences, Tsinghua University, Beijing 100084, China}
\author{W.~Cai}
\affiliation{Center for Quantum Information, Institute for Interdisciplinary Information
Sciences, Tsinghua University, Beijing 100084, China}
\author{X.~Mu}
\affiliation{Center for Quantum Information, Institute for Interdisciplinary Information
Sciences, Tsinghua University, Beijing 100084, China}
\author{W.~Dai}
\affiliation{Center for Quantum Information, Institute for Interdisciplinary Information
Sciences, Tsinghua University, Beijing 100084, China}
\author{W.~Wang}
\affiliation{Center for Quantum Information, Institute for Interdisciplinary Information
Sciences, Tsinghua University, Beijing 100084, China}
\author{L.~Hu}
\affiliation{Center for Quantum Information, Institute for Interdisciplinary Information
Sciences, Tsinghua University, Beijing 100084, China}
\author{X.~Li}
\affiliation{Center for Quantum Information, Institute for Interdisciplinary Information
Sciences, Tsinghua University, Beijing 100084, China}
\author{J.~Han}
\affiliation{Center for Quantum Information, Institute for Interdisciplinary Information
Sciences, Tsinghua University, Beijing 100084, China}
\author{H.~Wang}
\affiliation{Center for Quantum Information, Institute for Interdisciplinary Information
Sciences, Tsinghua University, Beijing 100084, China}
\author{Y.~P.~Song}
\affiliation{Center for Quantum Information, Institute for Interdisciplinary Information
Sciences, Tsinghua University, Beijing 100084, China}
\author{Zhen-Biao Yang}
\email{zbyang@fzu.edu.cn}
\affiliation{Fujian Key Laboratory of Quantum Information and Quantum Optics, College of Physics and Information Engineering, Fuzhou University, Fuzhou, Fujian 350108, China}
\author{Shi-Biao Zheng}
\email{t96034@fzu.edu.cn}
\affiliation{Fujian Key Laboratory of Quantum Information and Quantum Optics, College of Physics and Information Engineering, Fuzhou University, Fuzhou, Fujian 350108, China}
\author{L.~Sun}
\email{luyansun@tsinghua.edu.cn}
\affiliation{Center for Quantum Information, Institute for Interdisciplinary Information Sciences, Tsinghua University, Beijing 100084, China}


\begin{abstract}

To realize fault-tolerant quantum computing, it is necessary to store quantum information in logical qubits with error correction functions, realized by distributing a logical state among multiple physical qubits or by encoding it in the Hilbert space of a high-dimensional system. Quantum gate operations between these error-correctable logical qubits, which are essential for implementation of any practical quantum computational task, have not been experimentally demonstrated yet. Here we demonstrate a geometric method for realizing controlled-phase gates between two logical qubits encoded in photonic fields stored in cavities. The gates are realized by dispersively coupling an ancillary superconducting qubit to these cavities and driving it to make a cyclic evolution depending on the joint photonic state of the cavities, which produces a conditional geometric phase. We first realize phase gates for photonic qubits with the logical basis states encoded in two quasiorthogonal coherent states, which have important implications for continuous-variable-based quantum computation. Then we use this geometric method to implement a controlled-phase gate between two binomially encoded logical qubits, which have an error-correctable function.

\end{abstract}
\maketitle
\vskip 0.5cm

\narrowtext

Quantum computers process information in a way fundamentally different from their classical counterparts, where information is encoded in the state of a collection of quantum bits (qubits) and algorithms are carried out by performing a sequence of gates on these qubits~\cite{Nielsen}. Unlike classical bits, qubits are vulnerable to decoherence arising from coupling to the environment and noises of the control fields, which is one of the main obstacles to construct a large-scale quantum computer. To make a quantum computer function under decoherence effects, quantum information has to be stored in logical qubits, with which errors can be detected and corrected. In traditional quantum error correction (QEC) schemes, a logical qubit is redundantly encoded in multiple physical qubits~\cite{Fowler}. QEC based on these kind of encoding schemes has been demonstrated in various systems, including nuclear spins~\cite{Cory1998, Knill2001Benchmarking}, nitrogen-vacancy centers in diamond~\cite{Waldherr2014, Taminiau2014, Cramer2016}, photons~\cite{Yao2012Experimental}, trapped ions~\cite{Chiaverini2004, Schindler2011, Nigg2014}, and superconducting qubits~\cite{Reed2012Realization, Kelly2015State, Corcoles2015, Riste2015, Ming2019Experimental}. To run a quantum algorithm with these logical qubits, it is necessary to be capable of performing quantum gate operations between them, but which have not been demonstrated yet.

Error-correctable logical qubits can also be constructed by encoding the quantum information in the large Hilbert space of a harmonic oscillator, whose state can be controlled by using an ancillary qubit resonantly~\cite{Law1996Arbitrary, Ben2003Experimental, Hofheinz2009} or dispersively~\cite{LeghtasPRA2013, Vlastakis, Krastanov2015, Heeres2015, Wang2017} coupled to it. The Schr\"{o}dinger cat code~\cite{LeghtasPRL2013, Mirrahimi2014} and the binomial code~\cite{Michael2016} are paradigms of this approach, with each of which demonstrations of QEC have been reported in superconducting circuits~\cite{Ofek2016, Hu2018}, where an ancillary transmon qubit dispersively coupled to a three-dimensional cavity is used to detect and correct the photon loss of the multiphoton logical qubit stored in the cavity. With similar setups, universal single-qubit gate sets based on both encodings were realized by the gradient ascent pulse engineering (GRAPE) method~\cite{heeres2017implementing, Hu2018}. Recently, a quantum controlled-NOT gate between two asymmetrically encoded photonic qubits, respectively, stored in two cavities has been demonstrated~\cite{Rosenblum2018}. This gate was realized by encoding the codewords of the control qubit on the vacuum state and two-photon state, which form a logical space where errors due to photon loss cannot be corrected. Entangling gate operations between two error-correctable logical qubits still remain elusive.

\begin{figure}[tb]
\includegraphics{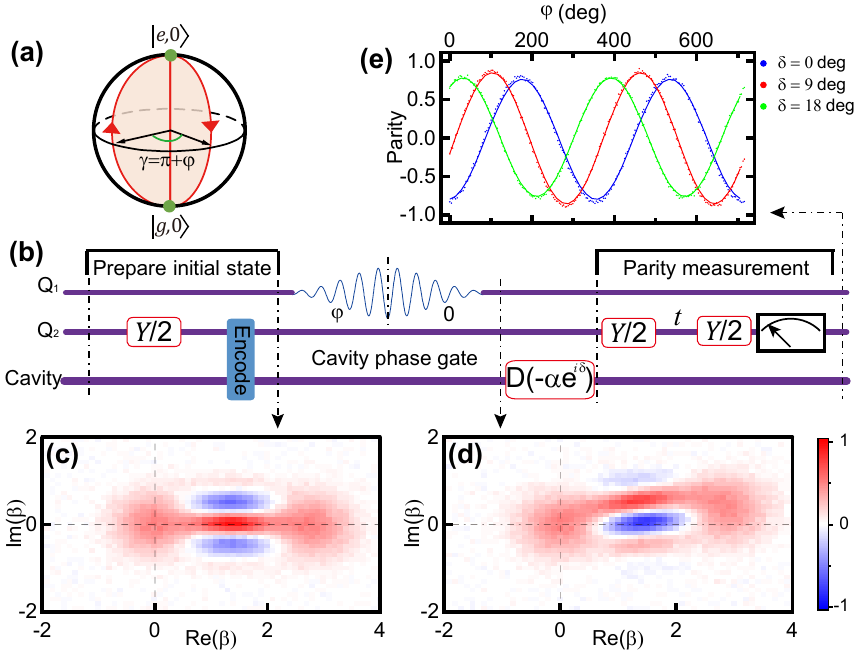}
\caption{Geometric manipulation of a photonic cat state. (a) Schematic of the nonadiabatic AA phase of a qubit. Two successive $\pi$ rotations of the qubit produce a geometric phase $\gamma = \pi + \varphi$, where $\varphi$ is the angle between the two rotation axes. (b) Experimental sequence to manipulate the cat state. A cavity is dispersively coupled to the qubit and initialized in a cat state $\left(\ket{0}+\ket{2\alpha}_c\right)/\sqrt{2}$ with the help of an ancillary qubit $Q_2$. The AA phase produced by the rotations of $Q_1$ conditional on the cavity's vacuum state is encoded in the probability amplitude of $\ket{0}$, resulting in a phase gate. (c) Measured Wigner function of the cavity state before the phase gate, corresponding to fidelity of 0.980 to the ideal cat state. (d) Wigner function of the cavity state after the gate with $\varphi=0$. The slight rotation and deformation of the Wigner function is due to the self-Kerr effect of the cavity. (e) Measured parity of the cavity state as a function of $\varphi$ after a displacement $D(-\alpha e^{i\delta} )$ for different values of $\delta$. Symbols are experimental data, in excellent agreement with numerical simulations~(solid lines).}
\label{fig:fig1}
\end{figure}

We here demonstrate a geometric method which enables realization of controlled-phase gates for photonic qubits with different encodings, in particular for two error-correctable logical qubits by using an ancillary transmon qubit dispersively coupled to the cavities storing the corresponding photonic qubits. With two successive carefully designed microwave pulses, the ancillary qubit is parallel transported along a closed loop on the Bloch sphere, picking up a geometric phase~\cite{berry1984quantal, Anandan1992The, Wilczek1989Geometric, Zheng2004Unconventional, Leek2007Observation, Song2017Geometric}, conditional on the particular component of the photonic qubits. The magnitude of the acquired geometric phase is controllable by the phase difference between the two applied pulses. We first employ this geometric phase to realize single- and two-cavity phase gates with coherent-state encoding. With this encoding, the single-cavity phase gate corresponds to manipulating the photon-number parity of a multiphotonic cat state. We further extend our method to implement a controlled-Z (CZ) gate between two binomial logical qubits, each of which has inherent error correction function. We demonstrate that this gate can evolve the two logical qubits to a maximally entangled state. The procedure can be straightforwardly and easily generalized to realize phase gates among multiple error-correctable logical qubits.

\begin{figure}[tb]
\includegraphics{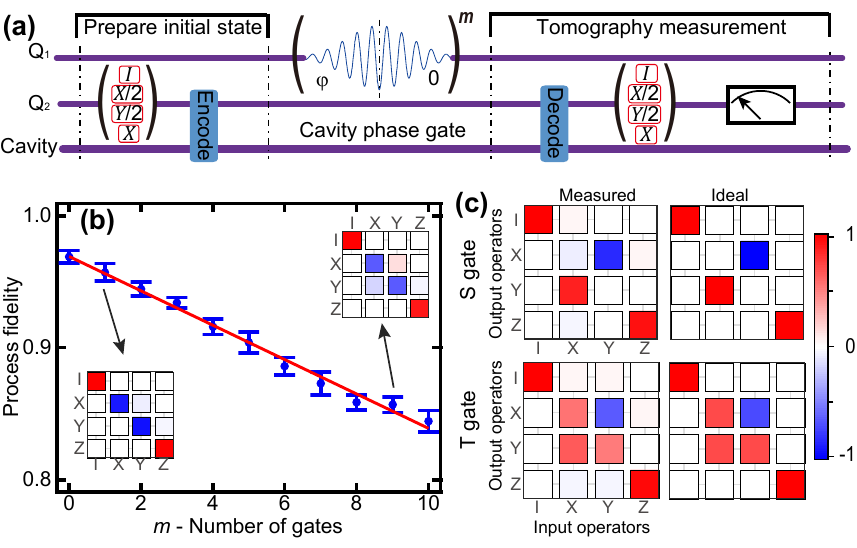}
\caption{Quantum process tomography~(QPT) of single-cavity geometric phase gates. (a) Experimental sequence. (b) The Pauli transfer process $R$ matrix fidelity as a function of $m$, the number of the Z gate on the cavity state. The insets show the measured $R$ matrices after one and nine Z gates, respectively. A linear fit of the process fidelity decay gives the Z gate fidelity $F_\mathrm{Z} = 0.987\pm0.001$. (c) The measured and ideal Pauli transfer $R$ matrices of the S gate and T gate with fidelities $F_{S} = 0.968$ and $F_{T} = 0.964$.}
\label{fig:fig2}
\end{figure}

\begin{figure}[tb]
\includegraphics{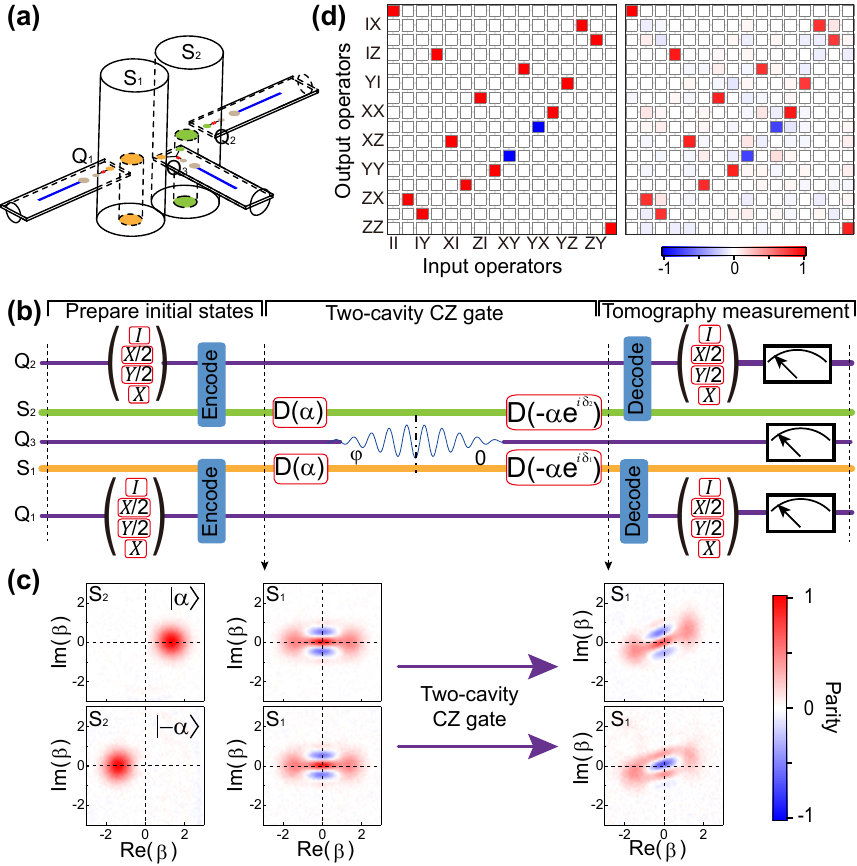}
\caption{Two-cavity geometric phase gate. (a) A 3D view of device B. A superconducting transmon qubit $Q_3$ at the center couples to two coaxial cavities $S_1$ and $S_2$, which couple to two other individual ancillary transmon qubits $Q_1$ and $Q_2$, respectively. Each of these transmon qubits independently couples to a stripline readout resonator used to perform simultaneous single-shot readout. (b) Schematic of the experimental sequence. (c) Measured individual Wigner functions of storage cavity $S_1$ and $S_2$. When the control cavity $S_2$ prepared in $\ket{\alpha}_c$~($\ket{-\alpha}_c$), the even cat state $\left( \ket{\alpha}_c + \ket{-\alpha}_c \right) /\sqrt{2}$ in target cavity $S_1$ evolves to even~(odd) cat state under the two-cavity CZ gate. The slight rotation and deformation of the Wigner functions after gate are due to the Kerr effect of the cavities. (d) Ideal~(left) and measured~(right) Pauli transfer $R$ matrices of the two-cavity CZ gate with the coherent encoding \{$\ket{0}_L = \ket{\alpha}_c$, $\ket{1}_L = \ket{-\alpha}_c$\}. The corresponding process fidelity $F_\mathrm{CZ\_ED}$~($F_\mathrm{ED}$) is 0.859~(0.954).}
\label{fig:fig3}
\end{figure}

The experiments presented in this work are based on two circuit quantum electrodynamics (QED) devices~\cite{Wallraff,Clarke2008Superconducting,You2011Atomic,devoret2013superconducting,Gu2017Microwave}. Device A, on which single-cavity geometric phase gates are performed, consists of two transmon qubits simultaneously dispersively coupled to two three-dimensional cavities~\cite{Paik, Kirchmair, Liu2017}. The parameters and architecture setup are described in Ref.~\cite{Xu2018}. Device B, on which two-cavity geometric phase gates are performed, consists of three transmon qubits dispersively coupled to two cylindrical cavities~\cite{Reagor2016} and three stripline readout cavities~\cite{axline2016an}. The device parameters are described in Ref.~\cite{Supplement}. In device A, the coupling between the qubit ($Q_1$) used to produce the geometric phase and the cavity used to encode this phase is described by the Hamiltonian
\begin{equation}
H=-\hbar \chi _{\mathrm{qs}}a^{\dagger }a\ket{e}\bra{e},
\end{equation}
where $\chi_{\mathrm{qs}}$ denotes the qubit frequency shift induced by per photon, $a^{\dagger }$ and $a$ are the creation and annihilation operators for the particular cavity field respectively, and $\ket{e}$ $(\ket{g})$ is the excited (ground) state of the qubit. In device B, the qubit, commonly coupled to two cavities used to store the photonic qubits, undergoes a frequency shift dependent on the photon numbers of both cavities.

The geometric manipulation technique is well exemplified with the even cat state $\left(\ket{\alpha}_c + \ket{-\alpha}_c \right)/\sqrt{2}$, where $\ket{\alpha}_c$ and $\ket{-\alpha}_c$ are coherent states, which can act as the two basis states of a logical qubit when  $_c\left\langle \alpha | -\alpha \right\rangle_c \approx O(e^{-2|\alpha|^2}) \ll 1$. To realize conditional qubit rotations, a phase-space displacement, $D(\alpha)$, is applied to the cavity, transforming its state to $\left(\ket{2\alpha}_c + \ket{0} \right)/\sqrt{2}$. The qubit, initially in the ground state $\left| g\right\rangle $, is then driven by a classical field on resonance with the qubit frequency conditioned on the cavity's vacuum state $\left| 0\right\rangle $. We here assume that the Rabi frequency $\varepsilon $ of the drive is much smaller than $\bar{n}\chi _\mathrm{qs}$, where $\bar{n} = 4|\alpha|^2$ is the average photon number of the state $\ket{2\alpha}_c$. In this case, the qubit's state is not changed by the drive when the cavity is in $\ket{2\alpha}_c$ due to the large detuning, and the system dynamics is described by the effective Hamiltonian
\begin{equation}
H_{\mathrm{eff}}=\frac 12\hbar \varepsilon e^{i\phi }\left| e\right\rangle
\left\langle g\right| \otimes \left| 0\right\rangle \left\langle 0\right|
+\mathrm{H.c.},
\end{equation}
where $\phi $ is the phase of the drive. This Hamiltonian produces a qubit rotation $R_{{\bf n}}^\theta $ conditional on the cavity's vacuum state, where $R_{{\bf n}}^\theta $ represents the operation that rotates the qubit's state by an angle $\theta =\int_0^\tau \varepsilon dt$ around the axis ${\bf n}$ with an angle $\phi $ to $x$ axis on the equatorial plane of the Bloch sphere, with $\tau $ being the pulse duration.

After two successive conditional $\pi $ rotations $R_{{\bf n}_1}^{\pi ,0}=R_{ {\bf n}_1}^\pi \otimes \left| 0\right\rangle \left\langle 0\right| $ and $R_{ {\bf n}_2}^{\pi ,0}=R_{{\bf n}_2}^\pi \otimes \left| 0\right\rangle \left\langle 0\right| $, the qubit makes a cyclic evolution, returning to the initial state $\left| g\right\rangle $ but acquiring a phase $\gamma =\pi+\Delta \phi = \Omega/2$, where $\Delta \phi =\phi _1-\phi _2$ represents the angle between the two rotation axes, and $\Omega$ is the solid angle subtended by the trajectory traversed by the qubit on the Bloch sphere, as shown in Fig.~\ref{fig:fig1}(a). This conditional phase shift leads to the cavity state $\left( \ket{2\alpha}_c + e^{i\gamma}\ket{0} \right)/\sqrt{2}$. A subsequent displacement $D(-\alpha )$ transforms the cavity to the state $(\ket{\alpha}_c +e^{i\gamma }\ket{-\alpha}_c )/\sqrt{2}$, realizing the phase gate. Because of the quantum interference of the two superposed coherent state components $\ket{\alpha}_c $ and $\ket{-\alpha}_c $, the cavity photon-number parity $P$ exhibits a periodical oscillation when the geometric phase $\gamma $ is varied: $P = \cos{\gamma}$. This procedure allows for manipulation of the parity of the cat state; when $\gamma=\pi$, the parity is reversed.

To simplify the operation, in our experiment the cavity displacement before the conditional qubit rotation is incorporated with the preparation of the initial cavity state; $\ket{2\alpha}_c$ and $\ket{0}$ instead act as the two logical basis states $\ket{0}_L$ and $\ket{1}_L$ for the single-cavity phase gate demonstration. We note that there is a compromise of choosing the value of $\alpha$. On one hand, a larger cat size is favorite for decreasing the overlapping between the two coherent state components, and for shortening the gate duration. On the other hand, the gate infidelity caused by the Kerr effects increases with the cat size. In our experiment, $\alpha = \sqrt{2}$; with this setting the total gate error is minimized. The experimental sequence to manipulate a cat state with device A is shown in Fig.~\ref{fig:fig1}(b). The cavity is initialized in the cat state $\left(\ket{2\alpha}_c + \ket{0} \right)/\sqrt{2}$ [the measured Wigner function is shown in Fig.~\ref{fig:fig1}(c)] with the help of ancillary qubit $Q_2$ following the GRAPE technique~\cite{Khaneja2005,deFouquieres2011}. The two subsequent conditional $\pi$ rotations on $Q_1$, yield a geometric phase  $\gamma = \pi + \varphi$ conditional on \ket{0}, where $\varphi$ is the angle between the two rotation axes. The Wigner function of the cavity state after this single-cavity geometric phase gate is shown in Fig.~\ref{fig:fig1}(d) with $\varphi=0$. After a displacement $D(-\alpha e^{i\delta} )$, the parity of the cavity state as a function of $\varphi$ is measured and shown in Fig.~\ref{fig:fig1}(e), in excellent agreement with numerical simulations.

Quantum process tomography~(QPT) is used to benchmark the cavity geometric phase gate performance, with the experimental sequence shown in Fig.~\ref{fig:fig2}(a). Since trusted operations and measurements necessary for QPT are unavailable in the coherent-state-encoded subspace, we characterize the gate by decoding the quantum information on the cavity back to the transmon qubit $Q_2$. We use the so-called Pauli transfer process $R$ matrix as a measure of our gate~\cite{Chow2012Universal}, which connects the input and output Pauli operators with $P_\mathrm{out} = RP_\mathrm{in}$. Figure~\ref{fig:fig2}(b) shows the $R$ matrix fidelity decay as a function of $m$, the number of the $\pi$ phase~(Z) gate. The fidelity at $m=0$ quantifies the ``round trip" process fidelity $F_{\mathrm{ED}}=0.969$ of the encoding and decoding processes only. A linear fit of the process fidelity decay gives the Z gate fidelity $F_Z = 0.987$, also consistent with the fidelity calculated from $F_Z = 1 - \left( F_\mathrm{ED} - F_{Z\_\mathrm{ED}} \right)$, where $F_\mathrm{Z\_ED}=0.957$ is the measured fidelity including the encoding and decoding processes. The measured and the ideal Pauli transfer $R$ matrices of the $S$ gate and $T$ gate are shown in Fig.~\ref{fig:fig2}(c), where $S = \ket{0}_L \bra{0} + i \ket{1}_L \bra{1}$ and $T =  \ket{0}_L\bra{0} + \exp{(i\pi/4)} \ket{1}_L \bra{1}$.

Our method can be directly generalized to implementation of controlled-phase gates between two photonic qubits encoded in two cavities that are dispersively coupled to one common superconducting qubit~\cite{Wang2016Schrodinger,Gao2018}. Figure~\ref{fig:fig3} shows the two-cavity geometric phase gates based on device B, whose schematic is shown in Fig.~\ref{fig:fig3}(a). Besides the transmon qubit commonly connected to both cavities, each cavity is individually coupled to another ancillary transmon qubit for encoding and decoding and measurement purposes. A two-cavity CZ gate with the coherent state encoding \{$\ket{0}_L = \ket{\alpha}_c$, $\ket{1}_L = \ket{-\alpha}_c$\} for both cavities is implemented by sandwiching a conditional qubit rotation between two pairs of displacement operations. The first pair of displacements transform the coherent states $\ket{\alpha}_c$ and $\ket{-\alpha}_c$ of each cavity to $\ket{2\alpha}_c$ and $\ket{0}$, respectively. The subsequent pulse, applied to the common qubit, produces a $2\pi$ rotation conditional on each cavity being in the vacuum state. The second pair of displacements restore each coherent state to the original amplitude. Consequently, the two cavities undergo a $\pi$ phase shift if and only if they are both in the logical state $\ket{1}_L$.

Here, we use the two-cavity QPT method to benchmark the performance of our realized CZ gate, with the experimental sequence shown in Fig~\ref{fig:fig3}(b). We first prepare the two cavities in a product state $\ket{0}_L \left( \ket{0}_L + \ket{1}_L \right)/\sqrt{2}$ or $\ket{1}_L \left( \ket{0}_L + \ket{1}_L \right)/\sqrt{2}$ in two separate experiments. After performing the two-cavity CZ gate, the even cat state $\left( \ket{\alpha}_c + \ket{-\alpha}_c \right)/\sqrt{2}$ in the target cavity $S_1$ evolves to even~(odd) cat state when the control cavity $S_2$ prepared in $\ket{0}_L$~($\ket{1}_L$), which is verified by the Wigner functions of the target cavity $S_1$ measured before and after the two-cavity CZ gate as shown in Fig~\ref{fig:fig3}(c).

With the two-cavity QPT method, we fully characterize the realized CZ gate with the measured Pauli transfer $R$ matrix, together with that for the ideal CZ gate, displaced in Fig.~\ref{fig:fig3}(d). The obtained process $R$ matrix fidelities, $F_{\mathrm{CZ\_ED}}$ and $F_\mathrm{ED}$, are respectively 0.859 and 0.954, which indicate the intrinsic two-cavity CZ gate fidelity is $F_{\mathrm{CZ}} = 0.905$, with the infidelities mainly coming from the control pulse imperfections~\cite{Supplement}.

\begin{figure}[tb]
\includegraphics{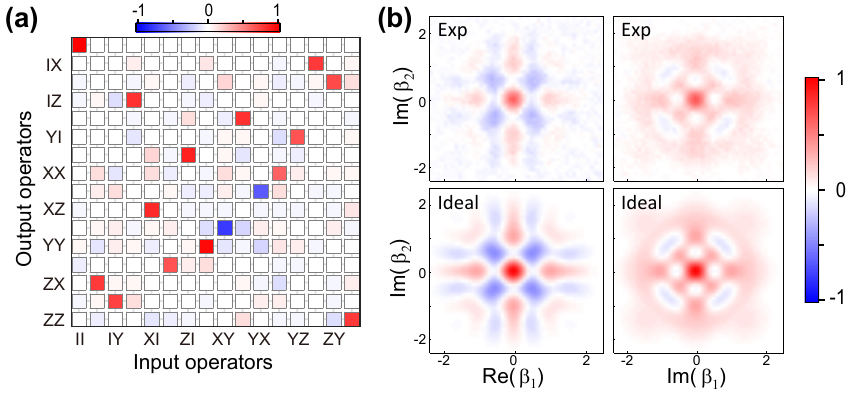}
\caption{Two-cavity CZ gate with binomial encoding. (a) Measured Pauli transfer $R$ matrix of the two-cavity CZ gate with the binomial encoding. The corresponding process fidelity $F_\mathrm{CZ\_ED}$~($F_\mathrm{ED}$) is 0.816~(0.922). (b) The measured and ideal joint Wigner function of the entangled logical Bell state $\ket{\Phi_+} = \left( \ket{01}_L + \ket{10}_L \right)/\sqrt{2}$ on the Im-Re and Im-Im planes, respectively.}
\label{fig:fig4}
\end{figure}

Our method allows implementation of a gate between two error-correctable logical qubits. For logical qubits whose basis states are encoded in even cat states, the photon-number parity can be used as an error syndrome of the single-photon loss~\cite{LeghtasPRL2013,Mirrahimi2014,Ofek2016, SunNature}. With this encoding, each of the two-qubit logical basis states is composed of four two-mode coherent state components, and a CZ gate can be realized by subsequently performing four conditional phase operations. We note that the displacements necessary for realizing these operations will move the logical qubits out of the error-correctable logical space. This problem can be overcome with another kind of error-correctable logical qubits binomially encoded as \{$\ket{0}_L=(\ket{0}+\ket{4}_F)/\sqrt{2}$, $\ket{1}_L=~\ket{2}_F$\}~\cite{Michael2016, Hu2018}.

To demonstrate the applicability of our method to binomial logical qubits, we first binomially encode the two cavities, then perform a CZ gate between thus-encoded qubits via geometric manipulation, and finally read out their joint state. The experimental sequence is similar to that in Fig.~\ref{fig:fig3}(b) but without the displacements. Because of the limitation of the dispersive couplings between the ancillary qubit and the cavities, the drive tuned to the ancilla's frequency associated with the cavities' basis state $\ket{22}_F$ will off-resonantly couple the ancilla's $\ket{g}$ and $\ket{e}$ states, and thus produce a small dynamical phase when the cavities are in other joint photon-number states. To minimize this dynamical effect and to speed up the gate, we successively apply two $\pi$ pulses to the ancilla: the first one has a duration of 20~ns and is nonselective; while the second one has a duration of 2~$\mu$s and involves nine frequency components, each selective on one of the following nine joint Fock states $\ket{j,k}_F$ ($j,k=0,2,4$). With suitable choice of the amplitudes and phases of these driving components, the resulting phase shift associated with the logical state $\ket{22}_F$ differs from those with other joint Fock states by $\pi$.

The two-cavity QPT method is also used here to benchmark the realized CZ gate with the binomial encoding, and the measured corresponding Pauli transfer $R$ matrix is displayed in Fig.~\ref{fig:fig4}(a). The corresponding process fidelity $F_\mathrm{CZ\_ED}$~($F_\mathrm{ED}$) obtained from the measured $R$ matrix is 0.816~(0.922), which indicates the intrinsic CZ gate fidelity $F_\mathrm{CZ} = 0.894$. We note that during the gate operation, it is unnecessary to change the photon numbers for both cavities, so that they remain in the original logical space. This gate, together with single-qubit rotations, allows generation of entangled Bell states for the two logical qubits, as shown in Fig.~\ref{fig:fig4}(b). We note that single-photon loss can be corrected with this encoding in principle, but the present gate is not realized fault tolerantly as the photon loss occurring during the gate will result in a random phase, destroying the stored quantum information. Recently, fault-tolerant phase gates on single binomially encoded photonic qubit were realized~\cite{reinhold2019errorcorrected, Ma2019Error}, however, fault-tolerant implementation of two-qubit gates remains an outstanding task.

Combined with additional single-cavity Hadamard gates of the binomial logical qubits realized by using the GRAPE technique, our two-cavity CZ gate can be used to directly generate an entangled logical Bell state $\ket{\Phi_+} = \left( \ket{01}_L + \ket{10}_L \right)/\sqrt{2}$. With the help of two ancillary qubits, joint Wigner tomography of the generated Bell state is performed. The upper row of Fig.~\ref{fig:fig4}(b) displays the two slice cuts of the measured two-mode Wigner functions for the generated Bell state, which agree well with those for the ideal logical Bell state shown in the lower row in Fig.~\ref{fig:fig4}(b). The fidelity of this entangled state, measured by decoding the logical states back to the ancillary qubits and then performing a joint state tomography, is 0.861.

Besides the controlled-phase gates, the geometric dynamics can be used to realize a two-cavity selective number-dependent arbitrary phase gate~\cite{Supplement}, which represents an extension of the previously reported selective number-dependent arbitrary phase operation for universal control of a single cavity state~\cite{Krastanov2015, Heeres2015}. The method can also be directly generalized to realize geometric gates among three or more cat-encoded or binomially encoded qubits by properly setting the driving pulse. This kind of gate is useful for quantum error correction~\cite{Reed2012Realization} and serves as a central element for implementation of the quantum search algorithm~\cite{Nielsen}.

\begin{acknowledgments}
We are grateful for valuable discussions with Chen Wang and Chang-Ling Zou. This work was supported by the National Key Research and Development Program of China No. 2017YFA0304303, the National Natural Science Foundation of China under Grants No. 11474177, No. 11874114, No. 11674060, No. 11875108, and No. 11874235, and the Natural Science Foundation of Fujian Province under Grants No. 2018J01412.
\end{acknowledgments}


%

\end{document}


\title{Supplementary Material for ``Demonstration of controlled-phase gates between two error-correctable photonic qubits"}

\author{Y.~Xu}
\thanks{These two authors contributed equally to this work.}
\affiliation{Center for Quantum Information, Institute for Interdisciplinary Information
Sciences, Tsinghua University, Beijing 100084, China}
\author{Y.~Ma}
\thanks{These two authors contributed equally to this work.}
\affiliation{Center for Quantum Information, Institute for Interdisciplinary Information
Sciences, Tsinghua University, Beijing 100084, China}
\author{W.~Cai}
\affiliation{Center for Quantum Information, Institute for Interdisciplinary Information
Sciences, Tsinghua University, Beijing 100084, China}
\author{X.~Mu}
\affiliation{Center for Quantum Information, Institute for Interdisciplinary Information
Sciences, Tsinghua University, Beijing 100084, China}
\author{W.~Dai}
\affiliation{Center for Quantum Information, Institute for Interdisciplinary Information
Sciences, Tsinghua University, Beijing 100084, China}
\author{W.~Wang}
\affiliation{Center for Quantum Information, Institute for Interdisciplinary Information
Sciences, Tsinghua University, Beijing 100084, China}
\author{L.~Hu}
\affiliation{Center for Quantum Information, Institute for Interdisciplinary Information
Sciences, Tsinghua University, Beijing 100084, China}
\author{X.~Li}
\affiliation{Center for Quantum Information, Institute for Interdisciplinary Information
Sciences, Tsinghua University, Beijing 100084, China}
\author{J.~Han}
\affiliation{Center for Quantum Information, Institute for Interdisciplinary Information
Sciences, Tsinghua University, Beijing 100084, China}
\author{H.~Wang}
\affiliation{Center for Quantum Information, Institute for Interdisciplinary Information
Sciences, Tsinghua University, Beijing 100084, China}
\author{Y.~P.~Song}
\affiliation{Center for Quantum Information, Institute for Interdisciplinary Information
Sciences, Tsinghua University, Beijing 100084, China}
\author{Zhen-Biao Yang}
\email{zbyang@fzu.edu.cn}
\affiliation{Fujian Key Laboratory of Quantum Information and Quantum Optics, College of Physics and Information Engineering, Fuzhou University, Fuzhou, Fujian 350108, China}
\author{Shi-Biao Zheng}
\email{t96034@fzu.edu.cn}
\affiliation{Fujian Key Laboratory of Quantum Information and Quantum Optics, College of Physics and Information Engineering, Fuzhou University, Fuzhou, Fujian 350108, China}
\author{L.~Sun}
\email{luyansun@tsinghua.edu.cn}
\affiliation{Center for Quantum Information, Institute for Interdisciplinary Information Sciences, Tsinghua University, Beijing 100084, China}

\pacs{} \maketitle

\section{Experimental Device and setup}
The single-cavity geometric phase gates are performed on Device A, which consists of two transmon qubits simultaneously dispersively coupled to two three-dimensional (3D) cavities~\cite{Paik,Vlastakis,Liu2017,Wang2017,Hu2018}. Details of the device parameters and architecture setup are described in Ref.~\cite{Xu2018}. Here we only describe Device B in detail.

\begin{figure}[b]
\includegraphics{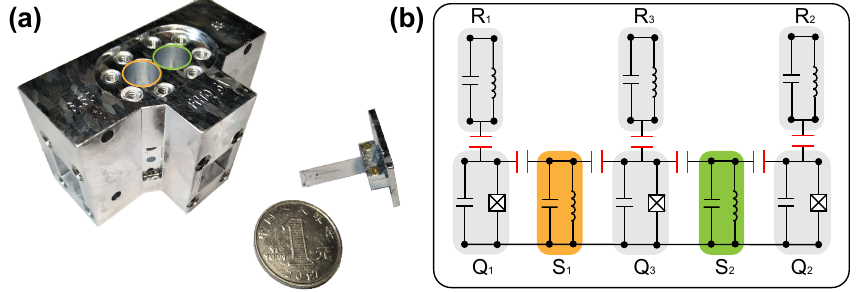}
\caption{(a) Optical image of Device B. The device is machined from a block of high-purity~(5N5) aluminum and is chemically etched. It consists of two coaxial stub cavities and three transmon qubits on different individual sapphire chips in three separate horizontal tunnels. Each qubit also couples to an individual stripline readout resonator. (b) Schematic of the effective circuit of Device B.}
\label{fig:Device}
\end{figure}

Device B with a circuit quantum electrodynamics (cQED) architecture contains two 3D coaxial stub cavities ($S_1$ and $S_2$), three superconducting transmon qubits ($Q_1$, $Q_2$, and  $Q_3$), and three stripline readout resonators ($R_1$, $R_2$, and  $R_3$). The 3D view of Device B is shown in Fig.~3(a) of the main text and the optical image is shown in Fig.~\ref{fig:Device}. The device is machined from a block of high-purity~(5N5) aluminum and is chemically etched to improve the surface quality~\cite{Reagor}. The two coaxial cavities are 3D $\lambda/4$ transmission line resonators~\cite{Reagor2016,Wang2016Schrodinger,Gao2018} with a center conductor of 3.3~mm in diameter and a cylindrical wall of 9.6~mm in diameter. The fundamental mode frequencies are mainly determined by the heights of the center stubs, 9.8~mm and 10.8~mm for $S_1$ and $S_2$ respectively. There are three horizontal tunnels housing three individual sapphire chips with patterned transmon qubits to couple to the two cavity modes. Qubit $Q_3$ on the middle chip is designed with three antenna pads to couple to the two coaxial cavity modes and one stripline readout mode, respectively. Each of the other two ancillary qubits ($Q_1$ and $Q_2$) only has two antenna pads to couple to the corresponding cavity mode and individual stripline readout resonator. Each stripline readout resonator is formed by the metal wall of the tunnel and an aluminum strip simultaneously patterned on the same chip with the transmon qubit through a standard double-angle evaporation process after a single electron-beam lithography step.

\begin{figure*}[htb]
\includegraphics{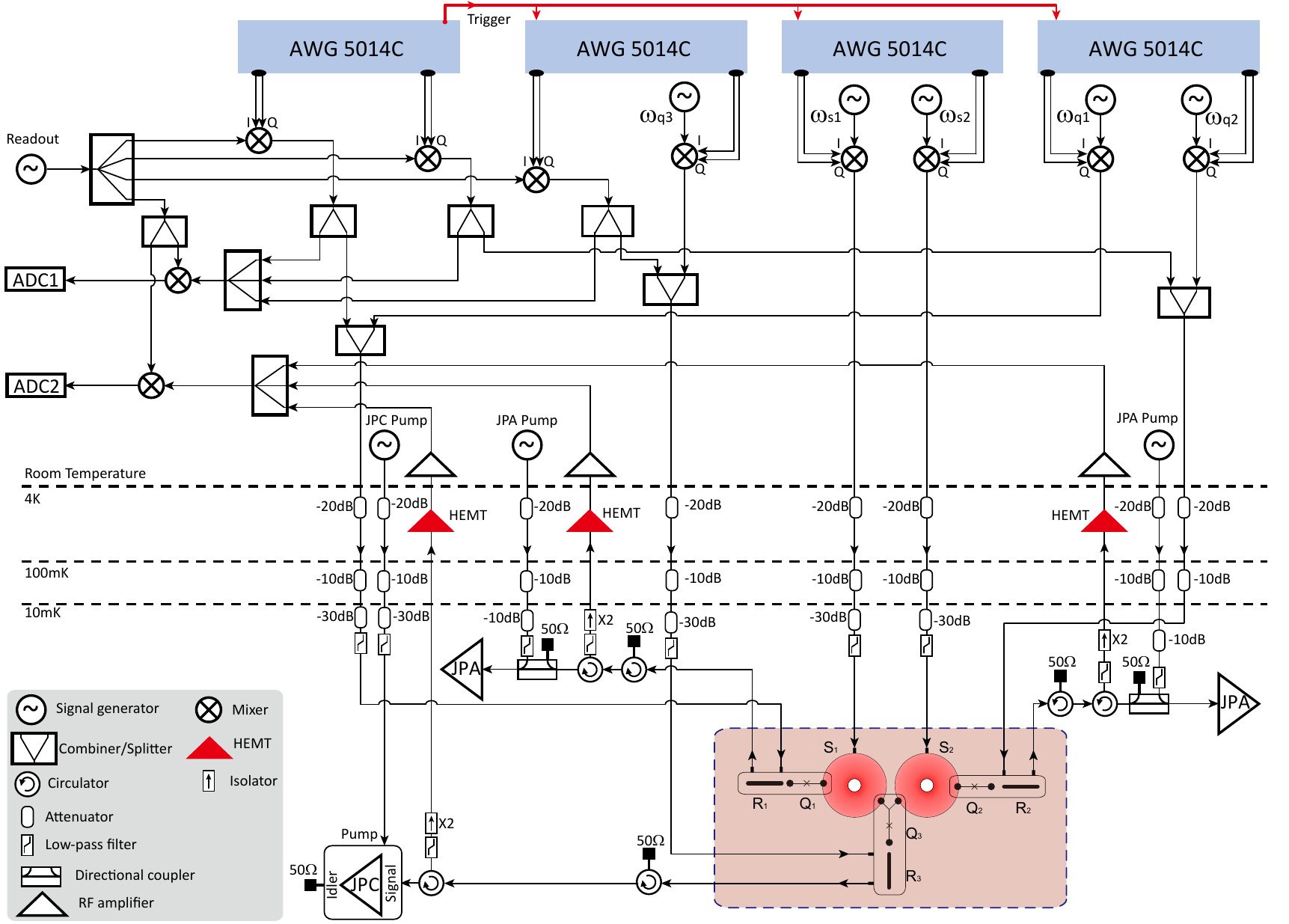}
\caption{Schematic of the full wiring of the experimental setup.}
\label{fig:setup}
\end{figure*}

The device is anchored to the mixing chamber of a cryogen-free dilution refrigerator which is cooled down to $T \approx 10$~mK. An additional magnetic shield covering the device is used to provide a clean electromagnetic environment. Attenuators and low-pass filters are used on the microwave lines to reduce the radiation noises of the signals. All the qubit and cavity drives are generated by IQ modulations with two analog channels of a Tektronix AWG5014C and an IQ mixer. The cavity states are initialized with the help of the ancillary qubits following the gradient ascent pulse engineering (GRAPE) technique~\cite{Khaneja2005,deFouquieres2011}. The qubit control pulses have a truncated Gaussian envelope with a width of $4\sigma=20$~ns. With the technique of ``derivative removal by adiabatic gate"~(DRAG) to remove the leakage and phase errors of the drive pulses~\cite{Motzoi,gambetta2011analytic}, the single qubit gates, characterized by the randomized benchmarking~(RB) method~\cite{knill2008randomized,chow2009randomized,magesan2012efficient,magesan2011scalable,barends2014superconducting}, result in an average fidelity of 0.9990, 0.9986, and 0.9992 for the three qubits respectively. The three qubits can be simultaneously measured with three individual readout control signals generated with different modulations of a same local oscillator~(LO). The readout signals are first amplified by quantum limited amplifiers at base temperature. We use two separate Josephson parametric amplifiers~(JPA) for $Q_1$ and $Q_2$, and a Josephson parametric converter~(JPC) for $Q_3$. Each readout signal is further amplified by a high electron mobility transistor~(HEMT) at 4K stage and a standard commercial RF amplifier at room temperature. Finally, the three readout signals are combined together and mixed down with the LO. After being digitized and recorded by the analog-to-digital converters~(ADC), the three readout signals can be distinguished through demodulations with different frequencies. The schematic of the full wiring of the experimental setup is shown in Fig.~\ref{fig:setup}.

\section{System Hamiltonian}
The three transmon qubits are dispersively coupled to the corresponding 3D cavity modes. Each transmon has a large anharmonicity and is considered as a two-level artificial atom, while each cavity mode is considered as a harmonic oscillator. Thus, the whole system can be described by the following Hamiltonian

\begin{eqnarray} \label{h}
\mathcal{H}/\hbar &=& \sum_{i=1}^{3}{\omega_{\mathrm{r}i} a_{\mathrm{r}i}^{\dagger}a_{\mathrm{r}i}}  + \sum_{i=1}^{3}\omega_{\mathrm{q}i}\ket{e_i}\bra{e_i} + \sum_{i=1}^{2}{\omega_{\mathrm{s}i}a_{\mathrm{s}i}^{\dagger}a_{\mathrm{s}i}} \notag\\
&-& \sum_{i=1}^{3}{\chi_{\mathrm{rq}i} \ket{e_i}\bra{e_i}a_{\mathrm{r}i}^{\dagger}a_{\mathrm{r}i}} \notag\\
&-& \chi_{\mathrm{s}1\mathrm{q}1} \ket{e_1}\bra{e_1}a_{\mathrm{s}1}^{\dagger}a_{\mathrm{s}1} - \chi_{\mathrm{s}1\mathrm{q}3} \ket{e_3}\bra{e_3}a_{\mathrm{s}1}^{\dagger}a_{\mathrm{s}1}\notag\\
&-& \chi_{\mathrm{s}2\mathrm{q}2} \ket{e_2}\bra{e_2}a_{\mathrm{s}2}^{\dagger}a_{\mathrm{s}2} - \chi_{\mathrm{s}2\mathrm{q}3} \ket{e_3}\bra{e_3}a_{\mathrm{s}2}^{\dagger}a_{\mathrm{s}2}\notag\\
&-& \sum_{i=1}^{2}{\frac{K_{\mathrm{s}i}}{2} a_{\mathrm{s}i}^{\dagger}a_{\mathrm{s}i}^{\dagger}a_{\mathrm{s}i}a_{\mathrm{s}i}} - \chi_{\mathrm{s}1\mathrm{s}2} a_{\mathrm{s}1}^{\dagger}a_{\mathrm{s}1}a_{\mathrm{s}2}^{\dagger}a_{\mathrm{s}2},
\end{eqnarray}
where $\omega_{\mathrm{r}i}$ is the readout resonator frequency of the $i$-th qubit with the corresponding ladder operators $a_{\mathrm{r}i}$ and $a_{\mathrm{r}i}^{\dagger}$; $\omega_{\mathrm{s}i}$ are the resonant frequency of the $i$-th storage cavity with the corresponding ladder operators $a_{\mathrm{s}i}$ and $a_{\mathrm{s}i}^{\dagger}$; $\omega_{\mathrm{q}i}$ is the transition frequency between the lowest two energy levels of the $i$-$\mathrm{th}$ qubit; $\chi_{\mathrm{rq}i}$ is the dispersive interaction between the $i$-th qubit and its corresponding readout resonator; $\chi_{\mathrm{s}1\mathrm{q}1}$, $\chi_{\mathrm{s}1\mathrm{q}3}$, $\chi_{\mathrm{s}2\mathrm{q}2}$, and $\chi_{\mathrm{s}2\mathrm{q}3}$ are the dispersive interactions between the three qubits and the two storage cavity modes; $K_{\mathrm{s}i}$ is the self-Kerr of the $i$-th storage cavity; and $\chi_{\mathrm{s}1\mathrm{s}2}$ is the cross-Kerr of the two storage cavities. All the relevant parameters in the Hamiltonian are experimentally measured and listed in Table~\ref{TableS1}.

\begin{table*}[htb]
\caption{Measured Hamiltonian parameters.}
\begin{tabular}{p{2.3cm}<{\centering} |p{2.3cm}<{\centering} |p{2.3cm}<{\centering} p{2.3cm}<{\centering} p{2.3cm}<{\centering} p{2.3cm}<{\centering} p{2.3cm}<{\centering} }
  \hline
   \multirow{2}{*}{Modes} &  \multirow{2}{*}{Frequency~(GHz)} & \multicolumn{5}{c}{Nonlinear terms: $\chi_{ij}/2\pi$~(MHz) }\\
                          &                                   & $S_1$  & $S_2$  & $Q_1$  &  $Q_2$  &  $Q_3$  \\\hline
  &&&&&&\\[-0.9em]
  $S_1$ & 6.594 & 0.005 & 0.004 & 1.599 & - & 0.524 \\
  &&&&&&\\[-0.9em]
  $S_2$ & 6.050 & 0.004 & 0.016 & - & 2.670 & 1.494 \\
  &&&&&&\\[-0.9em]
  $Q_1$ & 6.038 & 1.599 & - & 252 & - & - \\
  &&&&&&\\[-0.9em]
  $Q_2$ & 5.170 & - & 2.670 & - & 207 & - \\
  &&&&&&\\[-0.9em]
  $Q_3$ & 5.560 & 0.524 & 1.494 & - & - & 151  \\
  &&&&&&\\[-0.9em]
  $R_1$ & 8.892 & - & - & 2.0 & - & - \\
  &&&&&&\\[-0.9em]
  $R_2$ & 8.800 & - & - & - & 2.0 & - \\
  &&&&&&\\[-0.9em]
  $R_3$ & 9.032 & - & - & - & - & 1.5 \\
  \hline
\end{tabular}\vspace{-6pt}
\label{TableS1}
\end{table*}

The coherence properties of the qubits and the cavity modes are also experimentally characterized with the standard cQED measurements. In particular, the coherence times $T_1$ and $T_2^*$ of the storage cavities are measured through the relaxing of the Fock state $\ket{1}_F$ and the dephasing of the superposition state $\left( \ket{0} + \ket{1}_F \right)/\sqrt{2}$, respectively~\cite{Reagor2016}. Both initial states are generated with the selective number-dependent arbitrary phase~(SNAP) gates~\cite{Heeres2015}. All the results are listed in Table~\ref{TableS2}. We note that, in current device, the thermal populations of the qubits are not negligible~(about $0.01-0.03$) and are the dominant sources to limit $T_2^*$ of the cavities through the strong dispersive interaction.

\begin{table}[tb]
\caption{Coherence properties of the system.}
\begin{tabular}{p{1.7cm}<{\centering} |p{1.7cm}<{\centering} p{1.7cm}<{\centering} p{1.7cm}<{\centering}}
  \hline
  Modes & $T_1$ & $T_2^*$ & $T_2^\mathrm{Echo}$\\\hline
  &&\\[-0.9em]
  $S_1$ & 480~$\mu$s & 559~$\mu$s & -\\
  $S_2$ & 692~$\mu$s & 312~$\mu$s & -\\
  $Q_1$ & 35~$\mu$s & 25~$\mu$s  & 56.0~$\mu$s\\
  $Q_2$ & 20~$\mu$s & 12~$\mu$s  & 20~$\mu$s\\
  $Q_3$ & 25~$\mu$s & 25~$\mu$s  & 30~$\mu$s\\
  $R_1$ & 58~ns & -  & -\\
  $R_2$ & 55~ns & -  & -\\
  $R_3$ & 86~ns & - & -\\
  \hline
\end{tabular}
\label{TableS2}
\end{table}

\section{Simultaneous readout}
\begin{figure}[htb]
\includegraphics{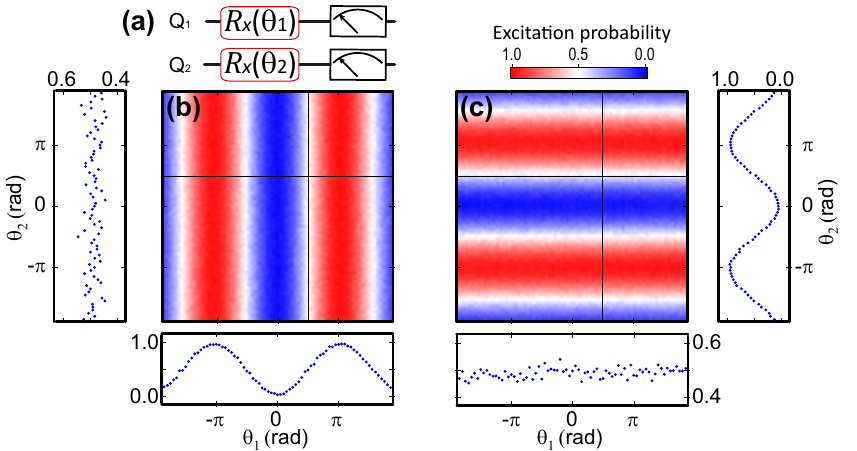}
\caption{Simultaneous Rabi experiments on the two ancillary qubits. (a) The experimental pulse sequence, where the two ancillary qubits are rotated along $x$ axis with independent angles $\theta_1$ and $\theta_2$, respectively, followed by simultaneous measurements on both of them. The measured excitation probabilities of qubits $Q_1$ and $Q_2$ as a function of the rotation angles $\theta_1$ and $\theta_2$ are shown in (b) and (c), respectively. The horizon and vertical cuts are also shown accordingly.}
\label{fig:TwoQubitRabi}
\end{figure}

In our experiment, each transmon qubit is connected to a quantum limited amplifier for fast high-fidelity single-shot readouts. The independence of the drives and measurements on them can be verified by simultaneous readouts of the two ancillary qubits, and the results are shown in Fig.~\ref{fig:TwoQubitRabi}.

In order to calibrate the three-qubit readout error, we prepare the system in each computational basis state and simultaneously measure the assignment probability $\vec{p} = \left( p_{000}, p_{001}, p_{010}, p_{011}, p_{100}, p_{101}, p_{110}, p_{111} \right)^T$ of the three qubits. By repeating the experiments for all the three-qubit computational basis states, we obtain the $8\times 8$ readout matrix $\mathcal{R}$ as shown in Table~\ref{TableS3}. We then can correct the readout errors by multiplying the inverse of the readout matrix $\mathcal{R}$ with the measured probability $\vec{p}$, such that $\vec{p}_\mathrm{corr} = \mathcal{R}^{-1}\cdot\vec{p}$ represents the real occupation probabilities of the eight computational basis states. For all the two-cavity experimental data shown in the main text, we have corrected the readout errors with this method.

\begin{table}[tb]
\caption{Three-qubit simultaneous readout assignment probability matrix $\mathcal{R}$. Each column represents the three-qubit measurement probabilities after preparing the qubits in the corresponding computational basis state.}
\begin{tabular}{p{0.8cm}<{\centering} |p{0.8cm}<{\centering} p{0.8cm}<{\centering} p{0.8cm}<{\centering} p{0.8cm}<{\centering} p{0.8cm}<{\centering} p{0.8cm}<{\centering} p{0.8cm}<{\centering} p{0.8cm}<{\centering}}
  \hline
   & $\ket{ggg}$ & $\ket{gge}$ & $\ket{geg}$ & $\ket{gee}$ & $\ket{egg}$ & $\ket{ege}$ & $\ket{eeg}$ & $\ket{eee}$ \\\hline
  &&&&&&&&\\[-0.9em]
  000 & 95.2 & 4.0 & 7.8 & 0.3 & 5.9 & 0.2 & 0.5 & 0.0 \\
  001 & 0.8 & 92.4 & 0.1 & 7.5 & 0.1 & 4.8 & 0.0 & 0.4 \\
  010 & 2.1 & 0.1 & 89.9 & 3.7 & 0.1 & 0.0 & 6.3 & 0.2 \\
  011 & 0.0 & 2.0 & 0.8 & 87.1 & 0.0 & 0.1 & 0.1 & 4.4 \\
  100 & 1.8 & 0.1 & 0.1 & 0.0 & 91.2 & 3.6 & 7.3 & 0.3 \\
  101 & 0.0 & 1.5 & 0.0 & 0.1 & 0.7 & 89.3 & 0.1 & 7.2 \\
  110 & 0.0 & 0.0 & 1.4 & 0.0 & 1.9 & 0.1 & 85.0 & 3.4 \\
  111 & 0.0 & 0.0 & 0.0 & 1.2 & 0.0 & 1.9 & 0.8 & 84.0 \\
  \hline
\end{tabular}
\label{TableS3}
\end{table}

\section{Encoding and decoding pulses}
In our experiment, in order to make the preparation and characterization of the cavity states easier, we employ optimal control pulses to encode and decode the cavity states, facilitated by the corresponding adjacent ancillary qubit. The encoding/decoding process corresponds to a unitary operation to realize a state mapping between the ancillary qubit and the storage cavity.

To realize the encoding/decoding of storage cavity $S_1$~($S_2$), we use the Hamiltonian defined in Eq.~\ref{h}~(only considering the relevant qubit-cavity modes), together with the control terms on the ancillary qubit $Q_1$~($Q_2$) and cavity $S_1$~($S_2$) in the form $\epsilon_{\mathrm{q}}(t) \sigma^+ + \epsilon_{\mathrm{q}}(t)^* \sigma^-$ and $\epsilon_{\mathrm{s}}(t) a_{\mathrm{s}}^{\dag} + \epsilon_{\mathrm{s}}(t)^* a_{\mathrm{s}}$, respectively. The temporal pulse envelopes $\epsilon_{\mathrm{q}}(t)$ and $\epsilon_{\mathrm{s}}(t)$ are discretized into 1~ns time step with piecewise constant, and numerically optimized with the quasi-Newton method to finally realize the target unitary $U_{\mathrm{tar}}$. 

In particular, for the encoding pulse, we wish to realize a unitary $U_{\mathrm{EN}}$ to perform the following state mapping:
\begin{eqnarray}
\left( c_0 \ket{g} + c_1 \ket{e} \right) \ket{0} \overset{U_{\mathrm{EN}}}{\longrightarrow} \ket{g} \left( c_0 \ket{0}_L + c_1 \ket{1}_L \right),
\end{eqnarray}
for all complex amplitudes $c_0$ and $c_1$ with both the coherent state encoding \{$\ket{0}_L = \ket{\alpha}_c$, $\ket{1}_L = \ket{-\alpha}_c$\} and the binomial encoding \{$\ket{0}_L=(\ket{0}+\ket{4}_F)/\sqrt{2}$, $\ket{1}_L=~\ket{2}_F$\}. This unitary process maps the quantum information stored in the ancillary qubit onto a superposition of logical basis states encoded in the cavity, while the ancillary qubit returns to the ground state after the operation.

For the decoding process, it simply reverses the above process: mapping the quantum information encoded in the cavity state in a superposition of logical basis states back onto the ancillary qubit state, while the cavity returns to the vacuum state after the operation. The decoding pulse is also used to eliminate the deterministic rotation and deformation due to the self-Kerr term in Eq.~\ref{h}. Therefore, after a geometric gate with a gate time $T$, the decoding pulse realizes the following state mapping:
\begin{eqnarray}
\ket{g} \left\{ e^{i\frac{K_{\mathrm{s}}}{2}a_s^\dag a_s^\dag a_sa_sT} \left( c_0 \ket{0}_L + c_1 \ket{1}_L \right) \right\} \overset{U_{\mathrm{DE}}}{\longrightarrow} \left( c_0 \ket{g} + c_1 \ket{e} \right) \ket{0},\notag\\
\end{eqnarray}
for all complex amplitudes $c_0$ and $c_1$. In order to make the decoding process more accurate, we have used the density matrices obtained from master equation simulations to perform the state mapping and calculate the decoding pulses.

In our implementation, the encoding/decoding pulses are first numerically calculated with the gradient descent method to realize the target state mapping and further optimized in the experiment in order to achieve the highest process fidelity.

\section{Quantum process tomography}
Both the single-cavity and two-cavity phase gates are characterized with full quantum process tomography~(QPT)~\cite{Nielsen}. We first prepare the cavity state with the encoding pulse. After performing the geometric gates on the cavity state, the decoding pulse is applied to map the cavity state back to the ancillary qubit state, which is reconstructed with pre-rotations \{$I$, $X_{\pi/2}$, $Y_{\pi/2}$, $X_{\pi}$ \} on each ancillary qubit before measurements. The density matrix is then reconstructed by the maximum likelihood estimation method~\cite{Daniel2001Measurement}.

Here, we use the Pauli transfer matrix $R$ to represent the quantum process, which is visually efficient and informative~\cite{Chow2012Universal}. In order to characterize the quantum process, before the encoding process we use initial states \{$\ket{g}$, $\ket{e}$, $\left( \ket{g} + \ket{e} \right)/\sqrt{2}$, $\left( \ket{g} -i \ket{e} \right)/\sqrt{2}$\} of the ancillary qubit for the single-cavity gates and \{$\ket{g}$, $\ket{e}$, $\left( \ket{g} + \ket{e} \right)/\sqrt{2}$, $\left( \ket{g} -i \ket{e} \right)/\sqrt{2}$\}$^{\otimes 2}$ of the ancillary qubits for the two-cavity gates. The $R$ matrix maps the input state vector $\vec{p}_\mathrm{in}$ with the output state vector $\vec{p}_\mathrm{out}$ by $\vec{p}_\mathrm{out} = R\cdot \vec{p}_\mathrm{in}$. The process fidelity is calculated from the measured $R$ matrix with
\begin{eqnarray}
F = \frac{\mathrm{Tr}\left( R^{\dag}R_\mathrm{ideal} \right)/d + 1}{d + 1},
\end{eqnarray}
where $R_\mathrm{ideal}$ is for a perfect process, $d = 2n$, and $n$ is the number of cavities.

\section{Joint Wigner tomography}
We verify the quantum entanglement of the two-cavity states with the joint Wigner tomography~\cite{Wang2016Schrodinger}, which is a measurement of the displaced joint photon number parity $P_J\left( \beta_1, \beta_2 \right)$ of the two cavities. We simultaneously map the parity of each cavity to its adjacent ancillary qubit. The single-shot readout of each qubit allows us to extract the displaced joint parity $P_J\left( \beta_1, \beta_2 \right)=P_1\left( \beta_1 \right)P_2\left( \beta_2 \right)$ by multiplying the two individually displaced cavity parities $P_1\left( \beta_1 \right)$ and $P_2\left( \beta_2 \right)$.

\begin{figure}[tbh]
\includegraphics{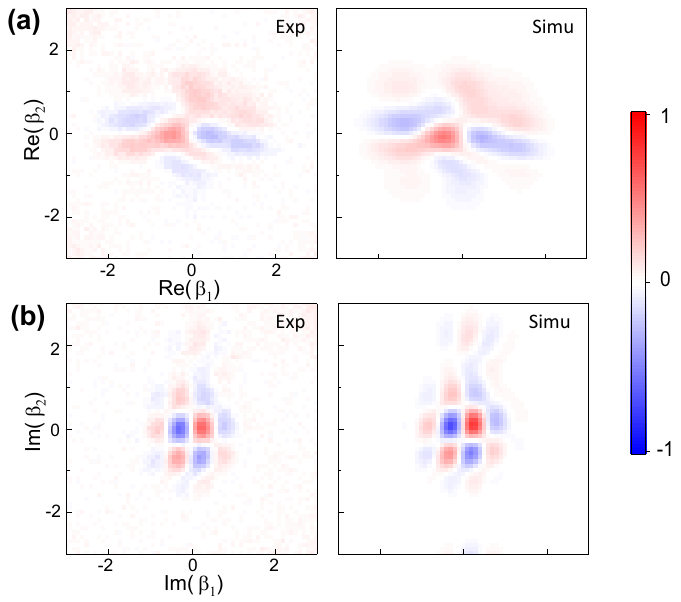}
\caption{Experimentally measured and simulated joint Wigner functions in the Re-Re plane~(a) and Im-Im plane~(b) for the entangled cavity states $\ket{\psi_1}$.}
\label{fig:JoinWigner}
\end{figure}

The realized two-cavity geometric controlled-Z (CZ) gate can be effectively used for generating entangled two-cavity states. For example of the coherent state encoded logical qubits, we first prepare the two storage cavities in a product state of two cat states $\ket{\psi_0} = \left( \ket{\alpha}_c + \ket{-\alpha}_c \right)\left( \ket{\alpha}_c + \ket{-\alpha}_c \right)/2$. After performing the two-cavity CZ gate, the two-cavity state evolves into an entangled state $\ket{\psi_1} = \left( \ket{\alpha}_c\ket{\alpha}_c + \ket{\alpha}_c\ket{-\alpha}_c + \ket{-\alpha}_c\ket{\alpha}_c -\ket{-\alpha}_c\ket{-\alpha}_c  \right)/2$ with a slight deformation due to Kerr effect. Figure~\ref{fig:JoinWigner} is the measured joint Wigner function of the entangled two-cavity state $\ket{\psi_1}$ in the Re-Re and Im-Im planes, which are consistent with our simulation results.

\begin{figure}[b]
\includegraphics{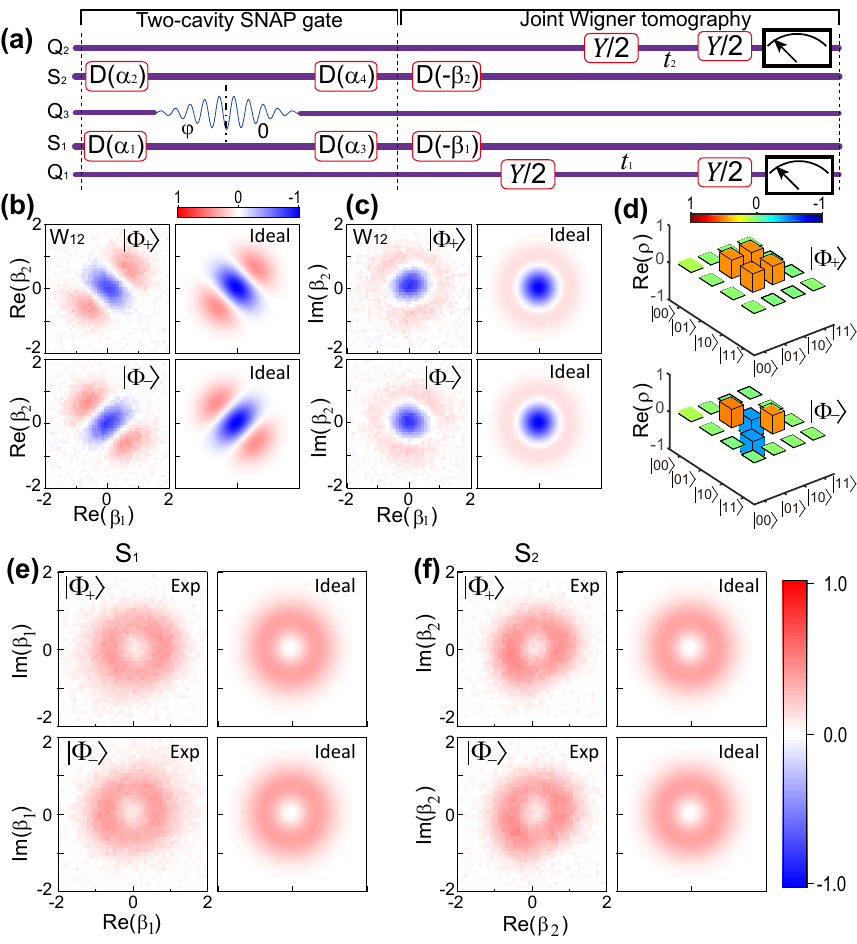}
\caption{Two-cavity SNAP gate to create single-photon Bell states $\ket{\Phi_{\pm}} = \left(\ket{01}_F\pm\ket{10}_F\right)/\sqrt{2}$. (a) Experimental sequence, which consists of a conditional $2\pi$ rotation on the qubit $Q_3$ and four displacements of the cavity states, followed by joint Wigner tomography measurements. (b) and (c) The measured joint Wigner function $W_{12}$ of the Bell states $\ket{\Phi_{+}}$ and $\ket{\Phi_{-}}$ on the Re-Re and Im-Re planes, respectively. (d) Real parts of the density matrices of the states $\ket{\Phi_{+}}$ and $\ket{\Phi_{-}}$ measured with the decoding and state tomography. Solid black outlines are for the ideal density matrices. Measured imaginary parts for both states are smaller than 0.03 and not shown. The fidelities for $\ket{\Phi_{\pm}}$ are 0.933 and 0.923, respectively. (e) and (f) Single-cavity Wigner functions of the entangled states $\ket{\Phi_{\pm}}$, measured from the storage cavities $S_1$ and $S_2$, respectively.}
\label{fig:twocavitySNAP}
\end{figure}

\begin{table*}[tbh]
\caption{Infidelities of the geometric gates. Error budgets for the single-cavity phase gate, two-cavity CZ gate with coherent encoding, and two-cavity CZ gate with binomial encoding. The conditional pulse selectivity error, the numerical optimization imperfection, and the self- and cross-Kerr induced error are considered as the control pulse imperfections.}
\begin{tabular}{p{6.5cm}<{\centering} |p{2cm}<{\centering} p{3.5cm}<{\centering} p{3.5cm}<{\centering}}
  \hline
  Error sources & Single-cavity phase gate & Two-cavity CZ gate with coherent encoding & Two-cavity CZ gate with binomial encoding\\\hline
  &&\\[-0.9em]
  Encoding/decoding error                    & 0.03            & 0.05           & 0.08\\
  Relaxation and dephasing                   & 0.01            & 0.02           & 0.04\\
  Conditional pulse selectivity error        & $<$0.01         & 0.06           & 0.03\\
  Numerical optimization imperfection        & -               & -              & 0.01\\
  Self- and cross-Kerr induced error         & $<$0.01         & 0.03           & $<$0.01\\
  \hline
  &&\\[-0.9em]
  Total                                      & 0.04        & 0.16           & 0.16\\
  \hline
\end{tabular}
\label{TableS4}
\end{table*}

\section{Two-cavity SNAP gates}
The geometric manipulation method can also be used to deterministically create high-fidelity single-photon Bell states $\ket{\Phi_{\pm}} = \left(\ket{01}_F\pm\ket{10}_F\right)/\sqrt{2}$, an extension of the previously reported SNAP operation for universal control of one cavity~\cite{Krastanov2015,Heeres2015} to two cavities. When combined with the single-cavity SNAP gates, our method can be used to realize arbitrary universal multi-cavity control. The experimental sequence is shown in Fig.~\ref{fig:twocavitySNAP}(a), where a conditional $2\pi$ rotation on qubit $Q_3$ is sandwiched in between two pairs of phase-space displacements of the cavities. The four displacement amplitudes, $\alpha_1 = \mp 0.8082$, $\alpha_2 = -0.8082$, $\alpha_3 = \pm 0.4103$, and $\alpha_4 = 0.4103$, are calculated from numerical simulation and further optimized in experiment in order to achieve higher state fidelities. With help of the two ancillary qubits, joint Wigner tomography of the two cavities is performed and two slice cuts of the measured two-mode Wigner function are shown in Figs.~\ref{fig:twocavitySNAP}(b-c). The density matrices of $\ket{\Phi_{+}}$ and $\ket{\Phi_{-}}$, reconstructed by mapping the state of the two cavities to qubits $Q_1$ and $Q_2$ and then jointly measuring the state of these qubits, are displayed in Figs.~\ref{fig:twocavitySNAP}(d), with state fidelities of 0.933 and 0.923, respectively.  Single-cavity Wigner functions on storage cavities $S_1$ and $S_2$ have also been performed and are shown in Fig.~\ref{fig:twocavitySNAP}(e-f). The measured single-cavity Wigner functions indicate mixed states of Fock states $\ket{0}$ and $\ket{1}_F$ as expected. Therefore, these results manifest that the generated two-cavity states are no longer separable and the individual cavity measurement destroys the coherence between them.

\section{Gate error analysis}
In this section, we estimate the main error sources and their contributions to the loss of fidelity for the single-cavity phase gate, two-cavity CZ gate with coherent encoding, and two-cavity CZ gate with binomial encoding. The results are summarized in Table~\ref{TableS4}.

1. The encoding/decoding error can be estimated by the measured process fidelity $F_\mathrm{ED}$ with the encoding and decoding processes only. The encoding/decoding errors mainly come from the decoherence of the ancillary qubit, imperfection of the GRAPE optimization, and inaccuracy of the Hamiltonian parameter calibration.

2. The relaxation and dephasing errors come from the ancillary qubit decoherence during the geometric gates with a gate time $T_\mathrm{gate}$. The ancillary qubit is in the excited state on average for half of the gate time during the selective pulse. Thus the average gate error induced by relaxation and dephasing can be estimated by $T_\mathrm{gate}/2T_1$. 

3. The selectivity error of the conditional qubit rotation, the numerical optimization imperfection, and the self- and cross-Kerr effects induced errors are considered as the control pulse imperfections and can be roughly estimated by the master equation simulation with only ideal operations. For the logical qubits with both coherent encoding and binomial encoding, the non-negligible self- and cross-Kerr terms in Hamiltonian Eq.~\ref{h} will result in deterministic and small deformation and rotation of the cavity states. In order to achieve high fidelity, we have partly compensated these deterministic effects by including them in the decoding pulses with numerical optimization.

The totally estimated infidelities are consistent with the experimentally measured gate process fidelities $F_\mathrm{Gate\_ED}$, which include the encoding and decoding processes.


%